\documentclass[3p,times,twocolumn]{elsarticle}

\usepackage{ecrc}
\usepackage[spanish,english]{babel}
\usepackage{subfigure}

\volume{00}

\firstpage{1}

\journalname{Nuclear and Particle Physics Proceedings}

\runauth{}

\jid{nppp}

\jnltitlelogo{Nuclear and Particle Physics Proceedings}

\usepackage{amssymb}

\usepackage[figuresright]{rotating}

\begin{document}

\begin{frontmatter}



\dochead{}

\title{Studies of high-energy pulsars: The special case of PSR J1849-0001}


\author[label1,label2]{Laila Vleeschower Calas}
\author[label1]{Sarah Kaufmann}
\author[label2]{C\'esar \'Alvarez Ochoa}
\author[label1]{Omar Tibolla}

\address[label1]{Mesoamerican Centre for Theoretical Physics, Universidad Aut\'onoma de Chiapas, Tuxtla Guti\'errez, M\'exico}
\address[label2]{Facultad de Ciencias en F\'isica y Matem\'aticas, Universidad Aut\'onoma de Chiapas, Tuxtla Guti\'errez, Mexico}

\begin{abstract}

We present the results from the data analysis of the XMM-Newton observation (53.6 ks) on PSR J1849-0001.
We studied in detail the X-ray emission of this pulsar and we found extended emission (up to $\approx 100 $ arcsec) from the Pulsar Wind Nebula (PWN), confirming that this is a case of a Pulsar/PWN system and strengthening the evidence that X-ray, hard X-ray and TeV gamma-ray sources are manifestations of the same system. Another important result of our study is the clear evidence that the X-ray PWN of PSR J1849-0001 is asymmetric.

\end{abstract}


\end{frontmatter}


\section{Introduction}
\label{Intro}

\vspace*{-0.15mm}
The number of very high energy gamma-ray emitting Galactic sources rapidly increased over the last years with the advent of the Imaging Atmospheric Cherenkov Telescopes and especially with the Galactic Plane Surveys by the H.E.S.S. experiment. Up-to-date VHE gamma-ray emission was found in many Pulsar Wind Nebulae (PWN) and Supernova Remnants (SNR) systems. 
But a very large part of the VHE gamma-ray emitters are still unidentified.

In the cases of PWN, the X-ray emitting region has generally a smaller extension than the TeV gamma-ray emission. 
Some PWN show symmetric X-ray emission, but there are famous examples of asymmetric X-ray emission, e.g. Crab, Vela and PSR B1509-58, representing the dynamics of this astrophysical objects. 

PSR J1849-0001 was detected with X-ray observations \cite{pulsar_1849} at the position of the soft gamma-ray source IGR J18490-0000 (detected by INTEGRAL) \cite{Terrier}. 
X-ray observations by RXTE revealed a pulsation period of 38.5 ms \cite{pulsar_1849}. 
Interestingly this source was also found to emit very high energy (VHE,  $E>10^{11}$ eV)  gamma-ray emission (discovered by the H.E.S.S. experiment) \cite{Terrier}. 
HESS J1849-000 is a very faint VHE source and hence only weak indication of its extension could be measured \cite{Terrier}. 

\subsection{Pulsar Characteristics}
\label{pulsar_charact}

The pulsar PSR J1849-0001 which is located at coordinates (J2000.0) R.A. = 18$^{h}$49$^{m}$0.1$^{s}$61, and decl. = -00$^{\circ}$01'17."6, is a 38.5 ms X-ray pulsar discovered in observations of the soft $\gamma$-ray source IGR J1849-0000 with the \textit{Rossi X-ray Timing Explorer (RXTE)}. 
This pulsar is spinning down rapidly with period derivative of 1.42 $\times$ 10$^{-14}$ s s$^{-1}$, yielding a spin-down luminosity of $\dot{E} = 9.8 \times 10^{36}$ erg s$^{-1}$, a characteristic age of $\tau_{c}=42.9$ kyr, and a surface dipole magnetic field of $B_{s} = 75 \times 10^{11}$ G \cite{pulsar_1849}.


\section{Observation Details}
\label{Obs_details}

The data analysed are from a 53.6 ks XMM-Newton observation of PSR J1849-0001 (ObsID 0651930201) that was acquired on March 23th, 2011, and using the European Photon Imaging Camera (EPIC) \cite{Turner}. 
The MOS cameras were operated in the Full Frame mode to have the full field of view (FOV) covered with a time resolution of 2.6 s.
 
The PN detector was operated in the Small Window mode. This mode 
 uses a smaller FOV
(63 x 64 pixels) to reduce the number of out-of-time events (1.1\%), to improve the time resolution to 5.7 ms, and, finally to reduce the pile-up effect for bright sources. 
Both instruments are sensitive to X-rays in the 0.2-12 keV range with energy resolution $\triangle E/E \approx 0.1 / \sqrt{E(keV)}$.


For the analysis we used the data already calibrated by the satellite (PPS, Processing Pipeline Subsystem).
The data were processed using the SAS version  15.0.0
pipeline, and were analysed using both the SAS and FTOOLS software packages. 
Only a small part of the observation was contaminated by flaring particle background, hence a good observing time of 47.9 ks for EPIC MOS and 49.2 ks for PN remained.

\begin{figure}
\hbox{
\centering
\includegraphics[width=0.45\linewidth]{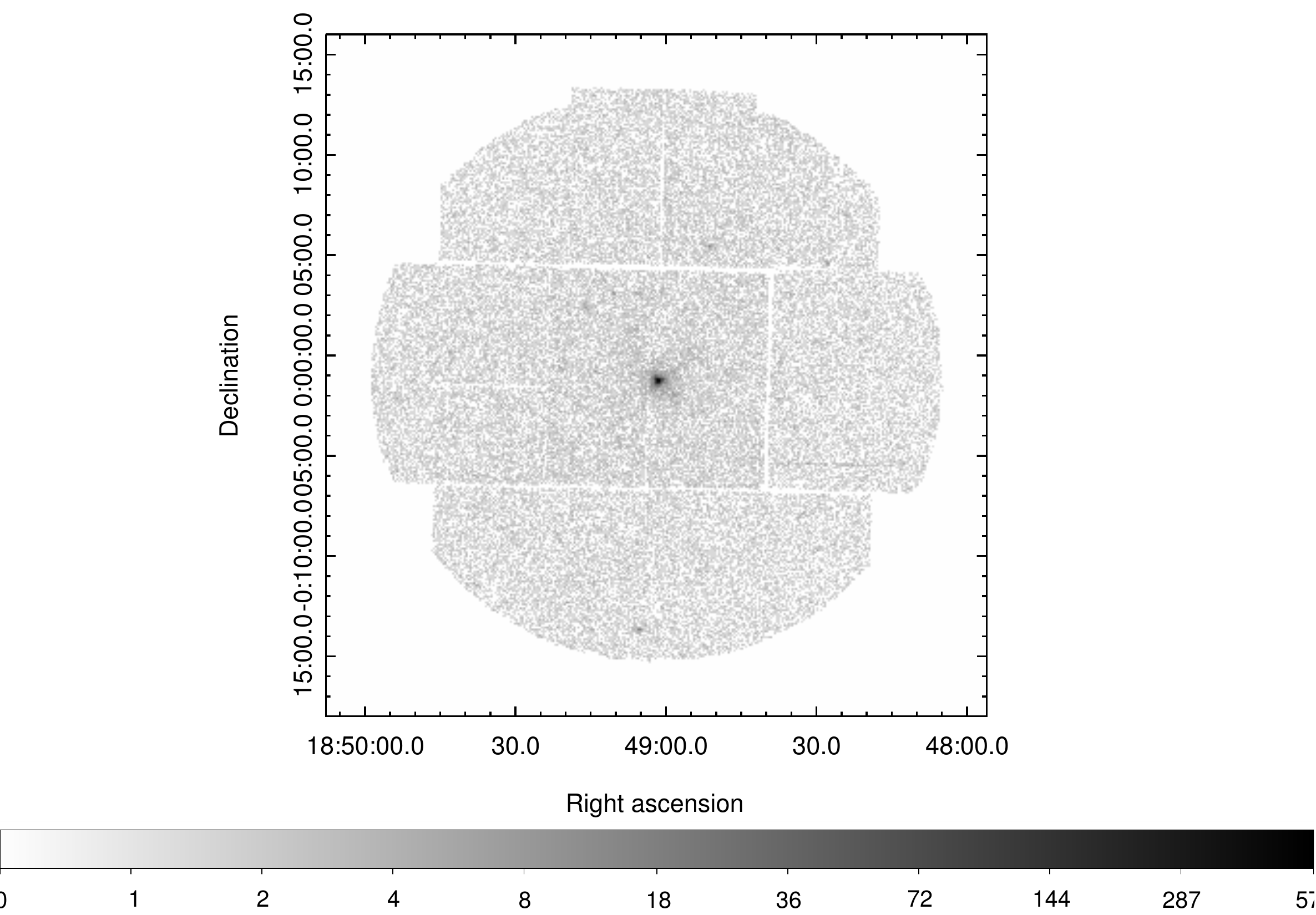}
\includegraphics[width=0.45\linewidth]{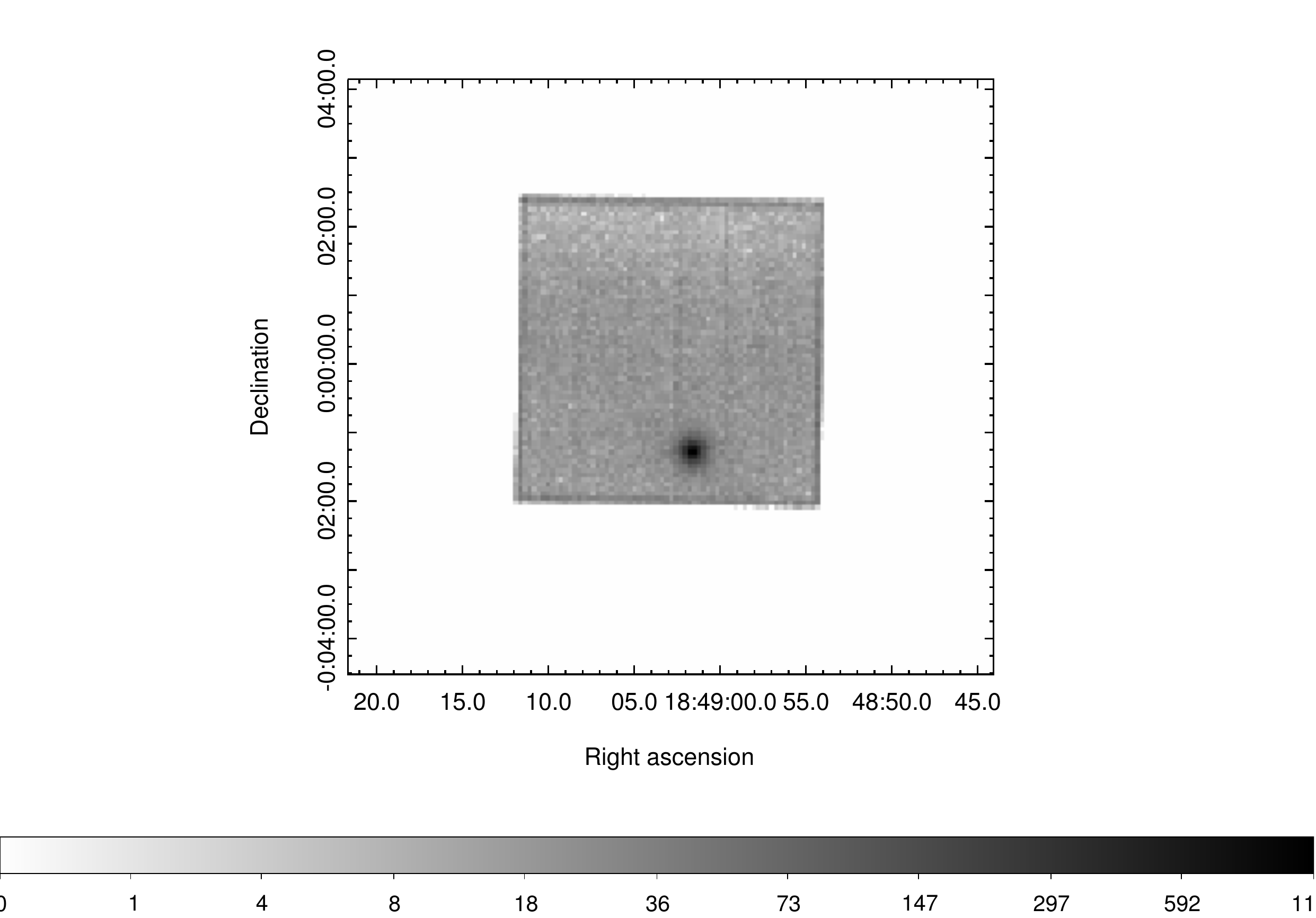}
}
\caption{\label{f:images_instruments} Images for MOS (left) and PN (right) cameras (filtered for flaring background periods).}
\end{figure}

As can be seen in Figure \ref{f:images_instruments}, we detected a strong X-ray point source at the position of PSR J1849-0001 and extended X-ray emission surrounding this point source which can be best seen in Figure \ref{f:image_2to7kev}. 

\section{Radial profile}

\begin{figure}
\centering
\includegraphics[width=65mm]{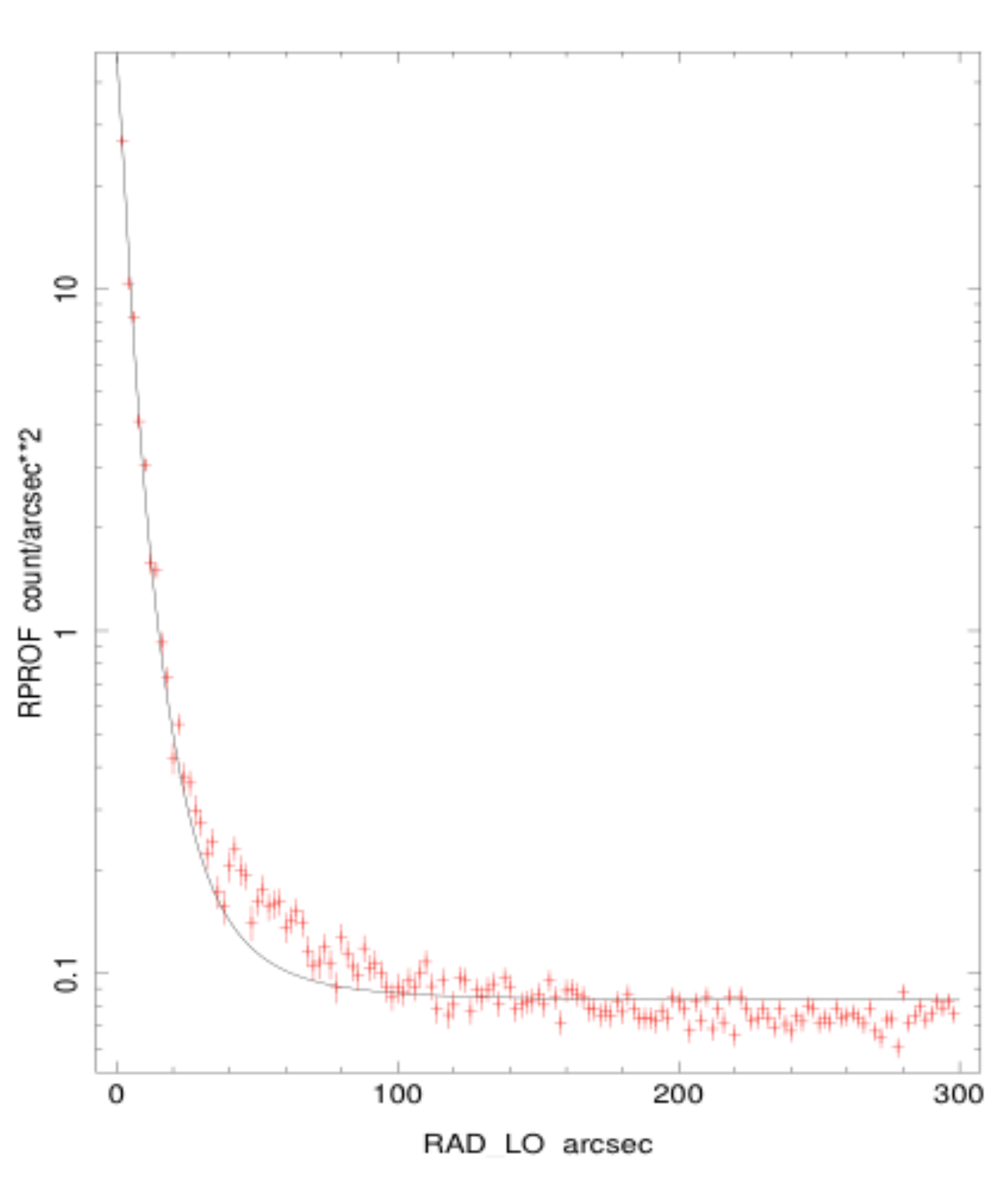}
\caption{Radial profile of the EPIC MOS1 detector}
\label{f:radial_profile_mos}
\end{figure}

To study the extended emission in more detail, we determined the radial profile in the EPIC MOS 1 and we fitted a model representing the point spread function (see Figure \ref{f:radial_profile_mos}). 
The extended emission is clearly visible up to approximately 100 arcseconds.

For the spectral analysis of the pulsar we used the region up to 20$"$ which is dominated by the pulsar. 
In the region 20$"$ to 40$"$, the amount of counts originating from the PSF wing of the pulsar can influence the spectrum of the extended emission and therefore we do not take this region into account for the spectral analysis. 
From 40$"$ to 100$"$ radius there is a clear excess of emission indicating that there is a region of extended emission. 
 Above 100$"$, where the radial profile reaches a constant, the emission is dominated by background photons.

\section{Spectral Analysis}

A spectrum of the point source and extended emission was extracted from each EPIC camera using the XMMSAS task \textit{evselect}, and were fitted using the XSPEC fitting package and using a single power-law model of the form $N(E) = N_0 \times E^{-\Gamma}$ for both EPIC PN and MOS instruments. 
The Galactic absorption in the direction of this source of $N_H = 1.57 \times 10^{22} cm^{-2}$ (LAB survey \footnote{http://heasarc.gsfc.nasa.gov/cgi-bin/Tools/w3nh/w3nh.pl}, \cite{Kalberla2005}).

\subsection{Point source spectra}

The spectra was extracted for each MOS camera using a circular region centred on the point source at a R.A. = 18$^{h}$49$^{m}$01$^{s}$.538, decl = -0$^{\circ}$01'17".22 with 20$"$ radius aperture, and a circular region of 60$"$ aperture placed on the right upper of the source region on the same CCD of each camera for the background located at R.A. = 18$^{h}$48$^{m}$47$^{s}$.538, decl. = +0$^{\circ}$02$'$15$"$.17 

In the case of PN, the regions were a circle with 20$"$ radius aperture centred on the point source and placed at R.A. = 18$^{h}$49$^{m}$01$^{s}$.538, decl = -0$^{\circ}$01$'$17$"$.22, and for the background an 40$"$ aperture at R.A. = 18$^{h}$48$^{m}$57$^{s}$.213, decl = +0$^{\circ}$01$'$30$"$.51.

\begin{figure}[h]
\centering
\subfigure[\label{f:mos1_spectra}]{\includegraphics[width=65mm]{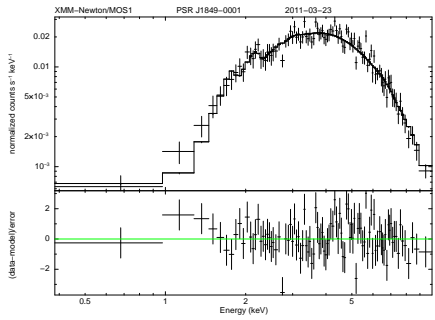}} 
\subfigure[\label{f:pn_spectra}]{\includegraphics[width=65mm]{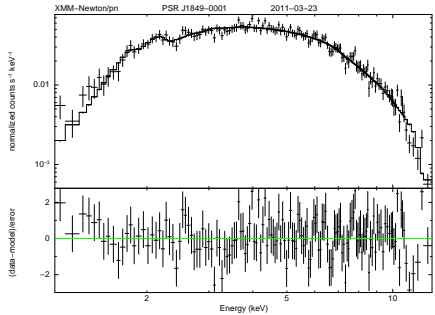}}
\caption{\label{f:source_spectra} XMM-Newton EPIC MOS1 and PN spectra (respectively) of the point source fitted to the model described in text.}
\end{figure}

\subsubsection{MOS 1}

The fit results in a good description of the data (see Figure \ref{f:mos1_spectra}) with a reduced $\chi^2 = 1.2$ for 95 degrees of freedom (dof). 
The spectrum has a photon index $\Gamma_{PSR}$ = 1.2 $\pm$ 0.1, an absorption of N$_{H}$ = (4.7 $\pm$ 0.45) $\times$ 10$^{22}$ cm$^{-2}$, and Normalization at 1 keV Norm. = (5.3 $\pm$ 1.3) $\times$ 10$^{-4}$ photons keV$^{-1}$ cm$^{-2}$ s$^{-1}$. 
The resulting flux is F$_{PSR}$ (2-10 keV) = (3.9 $\pm$ 0.2) $\times$ 10$^{-12}$ ergs cm$^{-2}$ s$^{-1}$ and the absorption corrected flux is F$_{PSR, unabs}$ (2-10 keV) = 5.0 $\times$ 10$^{-12}$ ergs cm$^{-2}$ s$^{-1}$. 
Results for MOS 2 are compatible.

\subsubsection{PN}

The data are good described with the fit resulting a reduced $\chi^{2}$ = 1.1 for 144 degrees of freedom (dof). 
The spectrum can be described by a photon index $\Gamma_{PSR}$ = 1.2 $\pm$ 0.1, an absorption of N$_{H}$ = (4.5 $\pm$ 0.3) $\times$ 10$^{22}$ cm$^{-2}$, and a Normalization at 1 keV = (5. $\pm$ 0.7) $\times$ 10$^{-4}$ photons keV$^{-1}$ cm$^{-2}$ s$^{-1}$. 
The resulting flux is F$_{PSR}$ (2-10 keV) = (3.9 $\pm$ 0.05) $\times$ 10$^{-12}$ ergs cm$^{-2}$ s$^{-1}$ and the absorption corrected flux is F$_{PSR, unabs}$ (2-10 keV) = (4.9 $\times$ 10$^{-12}$ ergs cm$^{-2}$ s$^{-1}$), see Figure \ref{f:pn_spectra}).

\begin{figure}
\centering
\includegraphics[width=55mm]{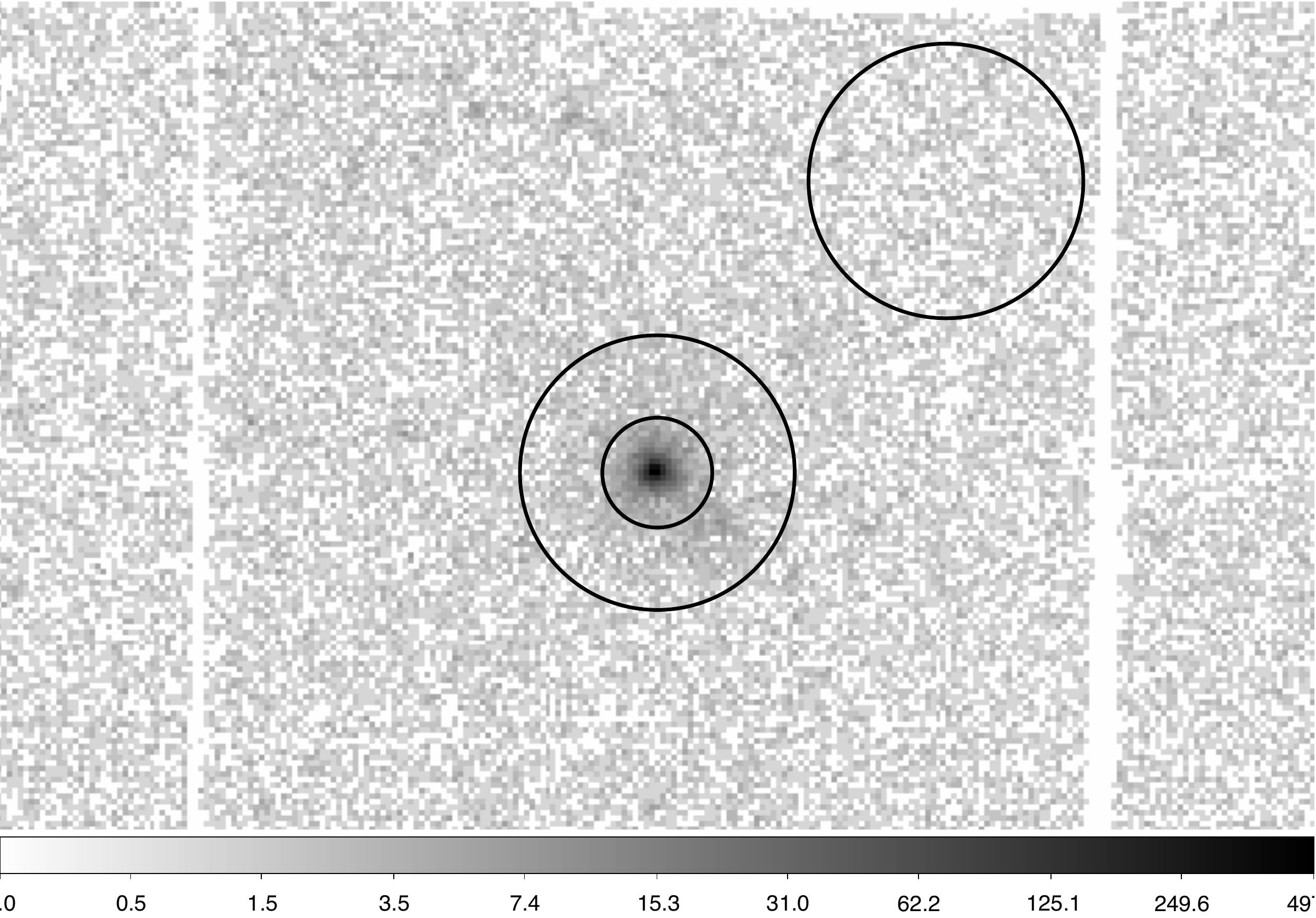}
\caption{Zoom showing better the extended emission. The annulus and background regions, which are used to estimate the spectra of the extended emission, are overplotted.}
\label{f:image_zoom_mos}
\end{figure}

\subsection{Extended X-ray Emission}

We extracted the spectrum of the extended nebula in an annulus surrounding the pulsar from 40" to 100" (see Figure \ref{f:image_zoom_mos}) radius at R.A. = 18$^{h}$49$^{m}$01$^{s}$.538, decl. = -0$^{\circ}$01$'$17$"$.22.

\subsubsection{MOS 1}

The resulting diffuse emission spectrum has $\Gamma_{PWN}$ = 1.7 $\pm$ 0.3, an absorption of $N_H$ = (5.08 $\pm$ 0.9) $\times 10^{22} \rm{cm^{-2}}$, and Normalization Norm. = (3.1 $\pm$ 4.0) $\times$ 10$^{-4}$ photons keV$^{-1}$ cm$^{-2}$ s$^{-1}$. 
The resulting flux in this case is $F_{PWN}$ (2-10 keV) = $(8.15 \pm 1.9) \times 10^{-13} \; \rm{erg \; cm^{-2} \; s^{-1}}$ and the absorption corrected flux is  $F_{PWN, unabs}$ (2-10 keV) = 1.158 $\times 10^{-12} \; \rm{ergs \; cm^{-2} \; s^{-1}}$. 
The reduced chi-squared is $\chi^2 = 1.0$ for 95 dof, (see Figure \ref{extended_em}).\\ The results for MOS 2 are also compatible. 
In this case, we did not extract the spectrum for EPIC PN detector, because our source is so close to the border of the small window (see Figure \ref{f:images_instruments}).

\begin{figure}
\centering
\includegraphics[width=65mm]{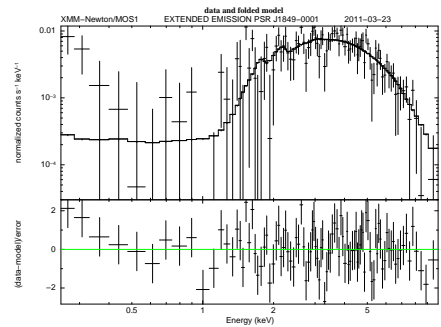}
\caption{XMM-Newton EPIC MOS 1 spectrum of the extended emission fitted to a power-law model (see the text).}
\label{extended_em}
\end{figure} 

\section{Morphology: the X-ray PWN is asymmetric}

\begin{figure}
\centering
\includegraphics[width=45mm]{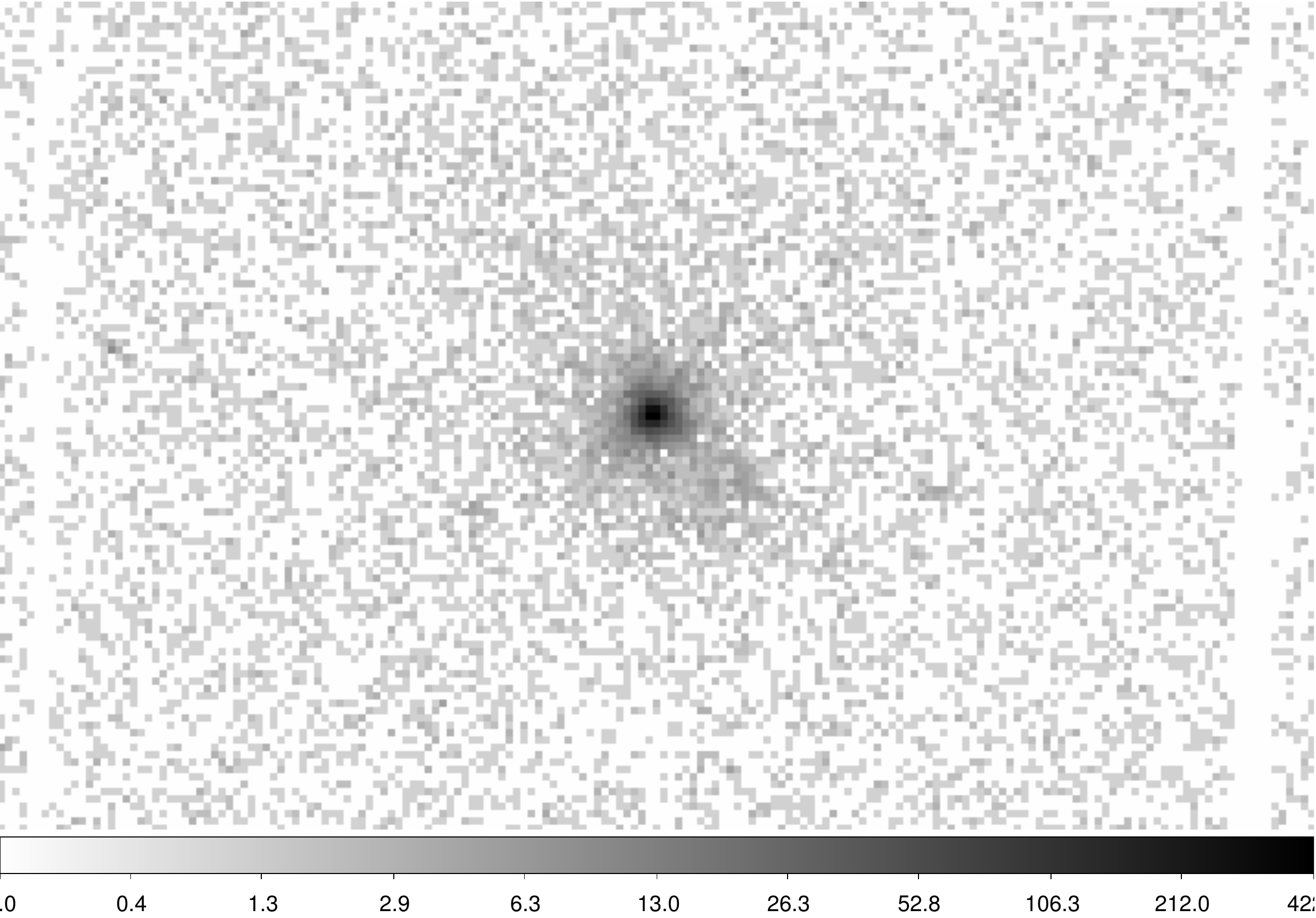}
\caption{Image of the system PSR-PWN in the energy range from 2 to 7 keV for MOS1.}
\label{f:image_2to7kev}
\end{figure}

 To study the morphology in more detail, an image from the energy range 2 to 7 keV was created. 
 In this energy range, the X-ray emission of the PWN is more significant, while at lower and higher energies the background becomes more dominant (Figure \ref{extended_em}). 
In Figure \ref{f:image_2to7kev} the extended X-ray emission can be seen very clearly.
Contours are applied and can be seen in figure \ref{extended_em}. The full extension of the PWN of around $100''$ and a clear asymmetric shape can be identified.


\begin{figure}
\centering
\includegraphics[width=0.9\linewidth]{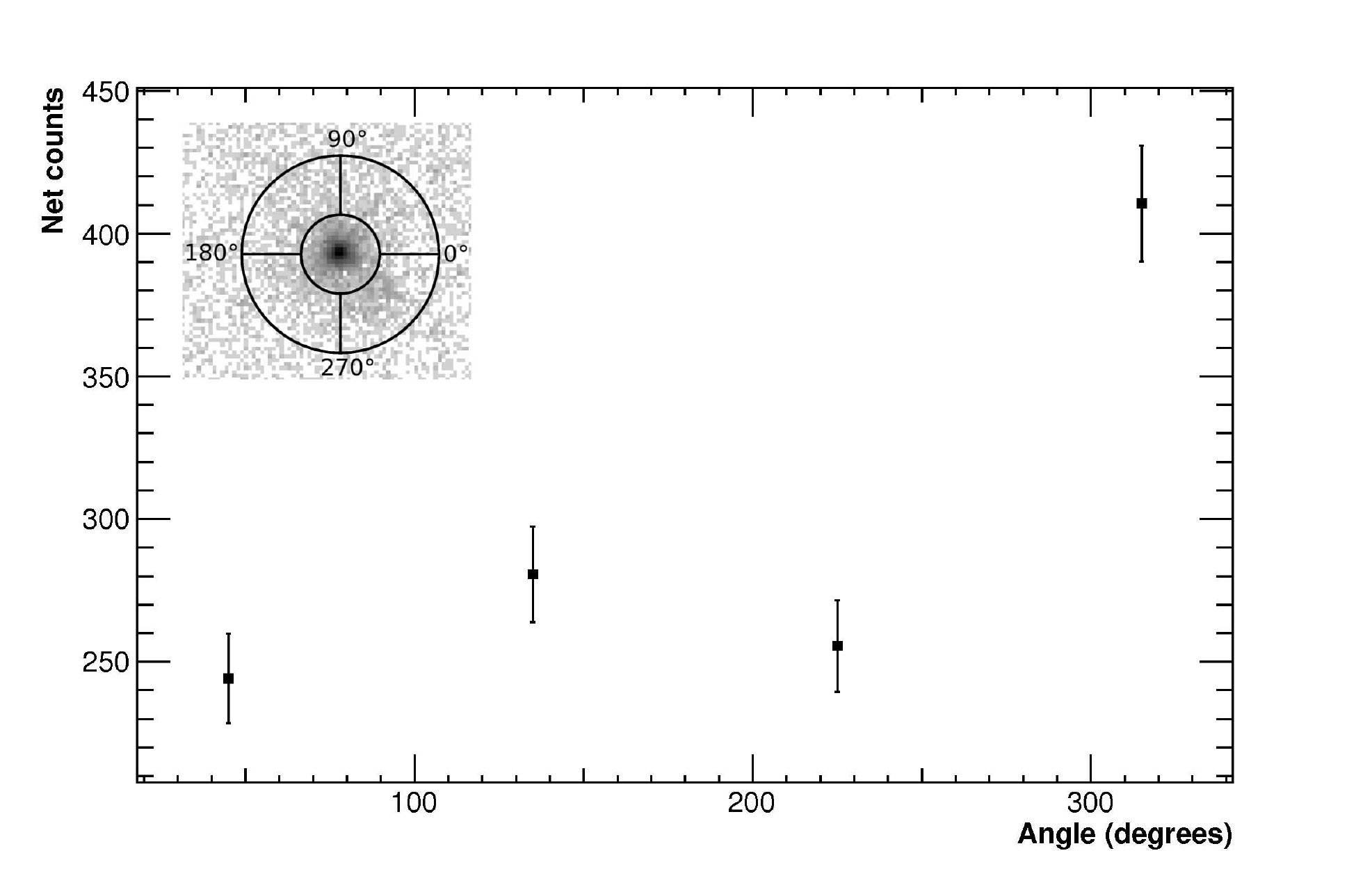}
\caption{\label{f:asymmetry}The number of counts of the PWN in four different sectors, ranging from 0 to 360 degrees, are determined (as illustrated in the inlay. A significant excess in the region from 270 to 360 degrees was found, confirming the asymmetry.}
\end{figure}

To probe that effectively there is an asymmetry on the extended emission, we calculate the number of counts in  four regions from 0 to 360 degrees (see Figure \ref{f:asymmetry}).
We used the annulus region ranging from 40$"$ to 100$"$, which is dominated by the extended emission (see radial profile) and which was used to determine the spectrum of the extended emission.
From Figure \ref{f:asymmetry} we can see clearly that there is a  significant excess on the region from 270 to 360 degrees confirming the asymmetry.

\begin{figure}
\centering
\includegraphics[width=55mm]{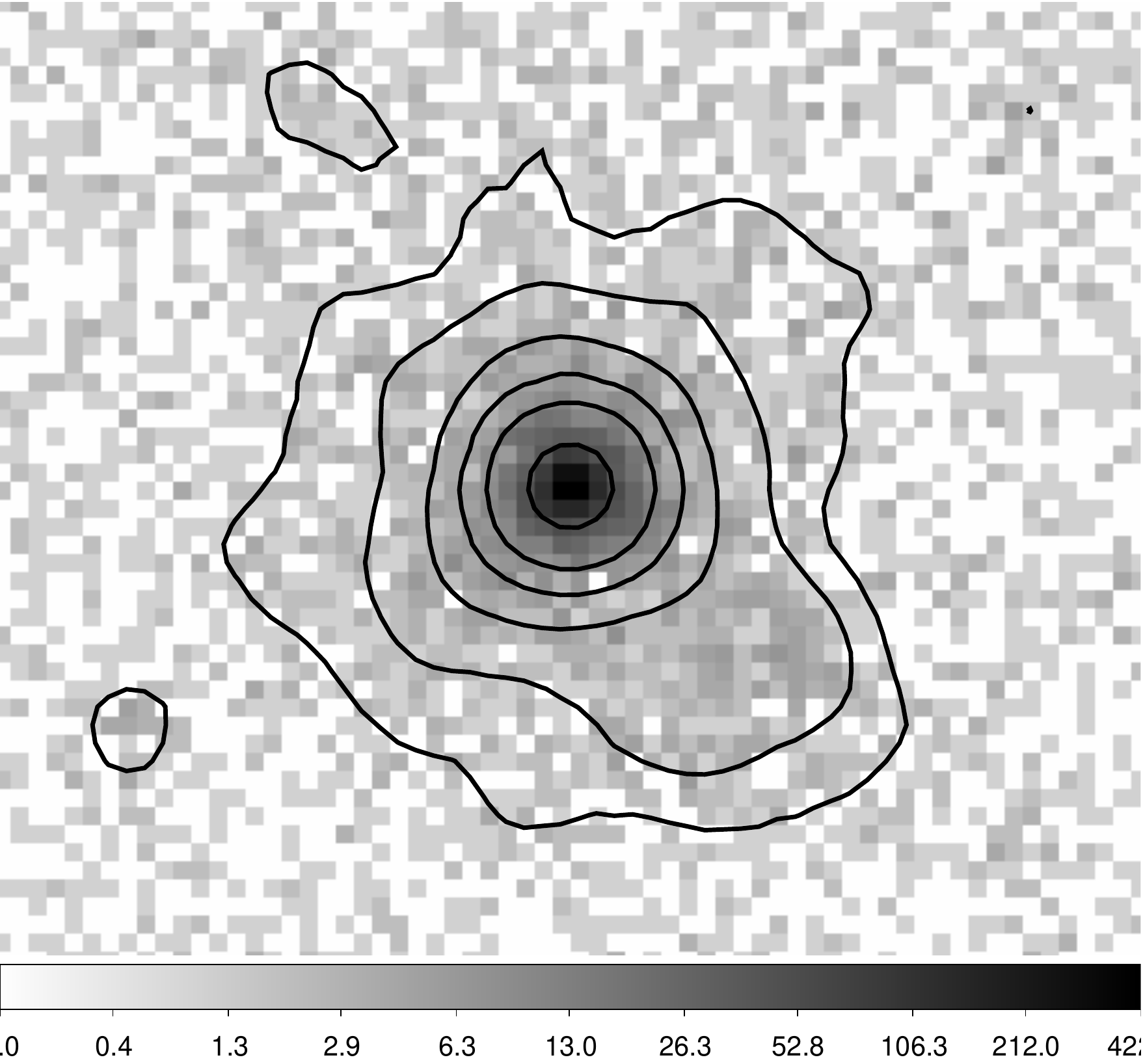}
\caption{Contours (representing similar flux level) are overplotted to the image of the energy range from 2 to 7 keV.}
\label{extended_em}
\end{figure}

\section{Summary}

The X-ray analysis of the long  exposure XMM-Newton observation at the position of the very high energy source IGR J18490-0000 reveals a bright X-ray pulsar and a clear detection of extended emission (up to $\approx$100 arcsec) interpreted as a pulsar wind nebula. Our studies confirm that this is a clear case of a Pulsar/PWN system and strengthen the evidence that the X-ray, hard X-ray and TeV gamma-ray sources are manifestations of the same system. 
The luminosity is consistent with the range of Pulsar/PWN systems \cite{Pavlov}. Modelling the results with \cite{Vorster},
seem to confirm that we deal with an intermediate age Pulsar/PWN system \cite{Kaufmann}. 

Moreover we found, that the derived hydrogen column density towards this source is very high. 
Though this source is in the Galactic plane, the measured column density is much larger than the average galactic hydrogen column density in this direction of the sky. 
This indicates that most of the absorbing material is very close to the source, probably circumstellar material.

 Another very important result is the asymmetry of the X-ray PWN. A significant excess is detected in the south-west region. 
 The X-ray PWN in this region is approximately 1.5 times brighter than the emission on the opposite side. 
 This could be interpreted as a movement of the pulsar over time resulting in a more luminous region where the pulsar passed.


\nocite{*}
\bibliographystyle{elsarticle-num}
\bibliography{martin}



\end{document}